\documentclass[ reprint, amsmath,amssymb, aps, nofootinbib]{revtex4-2}
\usepackage{graphicx}
\usepackage{dcolumn}
\usepackage{bm}
\usepackage{url}
\usepackage{color}
\begin{document}

\title{Thermodynamic Uncertainty Relation For a Multi-Phase Alternating Renewal Random Walk}
\author{Fatemeh Karimi}
\author{Farhad H. Jafarpour}
\email{f.jafarpour@basu.ac.ir}
\altaffiliation{Faculty of Science, Bu-Ali Sina University ,Hamedan, Iran}

%%%%%%%%%%%%%%%%%%%%%%%%%%%%%%%%%%%%%%%%%%%%%%%%%%%%%%%%%%%
\begin{abstract}
Thermodynamic Uncertainty Relations (TURs) impose universal bounds linking current precision to entropy production 
in nonequilibrium systems. While general bounds like the Proesmans--Van den Broeck (PV) relation provide a broad 
framework, they often remain loose for processes characterized by renewal events. In this work, we derive a generalized 
entropic bound for current fluctuations in renewal-reward processes. By utilizing a rigorous variance decomposition within 
the framework of renewal-reward theory, we obtain a model-independent bound that is rigorous and, for a range of
parameters in the examples considered, tighter than the PV relation evaluated with the total entropy production of the 
extended (position and heading) state space. Notably, in the linear-response regime, our bound correctly scales with the 
renewal rate and identifies a precision penalty arising from fluctuations of the cycle reward. These results provide a physically 
informative constraint on the precision of run-and-tumble-type dynamics and highlight the universal limits of transport in 
complex stochastic walkers.
\end{abstract}
%%%%%%%%%%%%%%%%%%%%%%%%%%%%%%%%%%%%%%%%%%%%%%%%%%%%%%%%%%%
\maketitle
%%%%%%%%%%%%%%%%%%%%%%%%%%%%%%%%%%%%%%%%%%%%%%%%%%%%%%%%%%%

\section{Introduction}\label{sec:intro}
The discovery of the Thermodynamic Uncertainty Relation (TUR) has fundamentally reshaped 
our understanding of nonequilibrium systems by establishing a universal trade-off between 
current precision and the entropy production required to sustain it 
\cite{Barato2015, Gingrich2016, Horowitz2020, Pietzonka2018}. Rooted in the large-deviation 
structure of nonequilibrium fluctuations \cite{MaesNetocny2008}, these relations are deeply 
connected to the broader framework of Macroscopic Fluctuation Theory \cite{Bertini2015}. 
While subsequent work has further generalized these bounds on current fluctuations and their 
thermodynamic costs \cite{PietzonkaBaratoSeifert2016}, the original formulations were primarily 
derived for time-homogeneous Markovian steady states. Such frameworks allow for the inference 
of energetic costs from macroscopic fluctuations, yet they often fall short when applied to complex 
biological and physical systems. Real-world processes--ranging from molecular motors driving 
intracellular transport to the run-and-tumble locomotion of chemotactic bacteria--frequently operate 
through discrete cycles of activity and reset, giving rise to non-Markovian dynamics that challenge 
standard TUR assumptions.

To address these systems, Shreshtha and Harris \cite{ShreshthaHarris2019} recently introduced a 
framework utilizing renewal-reward theory (RRT) \cite{Smith1958,Cox1962} to bound the uncertainty 
of run-and-tumble processes. While providing a significant step forward, their formulation relied on a 
specific microscopic scaling where rewards are accumulated as sums of independent and identically 
distributed (i.i.d.) increments. In biological reality, however, the internal dynamics of a ``run'' may be 
highly complex, involving temporal correlations, non-additive rewards, or intermittent resting phases 
that decouple the accumulation of current from the passage of time.

The looseness of standard TUR bounds in this setting has two distinct
physical origins, which we make explicit here. First, standard TURs bound
current precision by the \emph{total} entropy production of a
time-homogeneous Markov generator; when the reward accumulated within
a renewal cycle is the coarse-grained output of a high-dimensional internal
dynamics --- as in the conformational cycle of a molecular motor or the
flagellar-bundle reconfiguration underlying a bacterial run --- this
internal complexity can generate substantial dissipation that never
manifests as \emph{directional} current bias. The total entropy production
can then be large while the current precision
remains small, rendering the standard bound
essentially vacuous; this is the ``hidden entropy production'' phenomenon
familiar from the coarse-graining literature
\cite{KawaguchiNakayama2013,BoCelani2017}. Second, even in the absence of
hidden dissipation, a broad cycle-duration distribution inflates the current
variance through renewal-time statistics that have no
counterpart in the entropy production of the underlying jump process; a
bound derived for time-homogeneous continuous-time Markov chains has no
mechanism to capture this renewal-time contribution, and consequently
loosens as the duration distribution broadens.
In the present bound this effect enters only through the second moment
$\mathbb{E}[R^2]$ of the cycle reward, which grows with the fluctuations of the
run duration whenever the reward depends on it; the remaining contribution of
the duration statistics, the second term of the variance decomposition, is
discarded (see Sec.~\ref{sec:example1}).

In this work, we present a comprehensive generalization of the renewal-reward 
thermodynamic uncertainty relation. By treating the renewal cycle as a ``black box'', we 
derive a bound that is independent of the internal microscopic scaling of the rewards. 
Our approach utilizes a rigorous variance decomposition to isolate the fundamental 
current fluctuations from the cycle-level correlations between duration and displacement 
rather than other protocols \cite{VanVu2020}. This results in a universal bound that requires 
only the first two moments of the cycle-integrated rewards and durations. A key feature of our 
derivation is the use of an auxiliary sign process $X$, which encapsulates the directional bias 
established at each renewal event. This construction allows us to map the precision of the 
complex integrated current onto the entropy production $\bar{s}_X$ of 
the directional choice itself--the tumble, with the orientation taken to be odd
under time reversal (see Sec.~\ref{sec:general_result})--which
often serves as the fundamental broken symmetry driving active transport.

We focus on free run-and-tumble-type dynamics, i.e.\ force-free motion in a
homogeneous, isotropic medium, because only in this regime do they realize the
required i.i.d.\ cycle and reset structure, with a direction that is independent
of the run statistics; in the presence of external potentials, confinement or
environmental gradients these properties are generally lost, and we return to
the scope of this assumption in the last section.

Our results rest on the following assumptions, which we state explicitly at the
outset. (A1)~\emph{Renewal structure:} the cycle duration and the cycle-integrated
reward are i.i.d.\ across renewals, with finite first and second moments; within
a single cycle they may be arbitrarily correlated. 
(A2)~\emph{Independent direction:} the auxiliary sign $X$ is i.i.d.\ across
cycles, takes both values $\pm1$ with non-zero probability (microscopic
reversibility, so that $\bar s_X$ is finite), and is independent of both the
duration and the reward of the cycle.
(A3)~\emph{Odd orientation (interpretive):} if $X$ is odd under time reversal,
$\bar s_X$ is the entropy production of the biased walk generated by the
successive directional choices (Sec.~\ref{sec:general_result}). This assumption
is not needed for the validity of the bound, since the PV inequality for a
binary variable holds unconditionally (Appendix~\ref{app:PV_binary}); it is
needed only to interpret $\bar s_X$ as an entropy production. These assumptions are satisfied by 
free run-and-tumble motion, but are generally violated in the presence of 
external potentials, confinement or environmental gradients; we
return to this point in the last section.

We first analyze this general bound in the linear-response regime, corresponding to the limit of a weak 
directional bias. In this limit, we show that the precision recovers the standard linear-response TUR 
scaling, but is corrected by a dimensionless renewal penalty. This factor explicitly demonstrates 
how fluctuations of the cycle reward, including those induced by fluctuating run durations, suppress 
the achievable precision of a stochastic walker, even in the most efficient transport regimes.

To demonstrate the robustness of our framework, we provide two distinct examples that go beyond 
previous models. First, we consider a two-phase run--rest renewal process where a particle alternates 
between active biased motion and passive waiting states. We show that our bound remains analytically 
tractable and, in a broad parameter regime, is tighter than the PV bound evaluated with the total entropy
production of the extended state space (heading even under time reversal).
The two bounds involve different entropy productions, and neither implies the
other (Sec.~\ref{sec:example1}). Second, we investigate 
a non-Markovian run process where the internal increments follow an AR(1) autoregressive structure  \cite{BoxJenkins}. 
This ``stress test'' proves that our bound remains valid even in the presence of strong temporal memory 
and non-linear variance growth, where standard IID assumptions fail.

Collectively, these results provide a physically informative and mathematically rigorous framework for bounding the 
precision of complex stochastic processes. By shifting the focus from microscopic increments to cycle-level 
statistics, our work offers a model-independent tool for characterizing the universal limits of transport in 
both biological and synthetic active matter systems.

This paper is organized as follows: in section \ref{sec:general_result} by considering a general renewal 
process we present a new TUR bound.  In section
\ref{sec:small_bias} the linear-response theory is checked for the bound. In sections \ref{sec:example1} 
and \ref{sec:example2} two examples from different classes of renewal processes are provided and 
studied. The last section is devoted to the summary of results and outlook. 
%%%%%%%%%%%%%%%%%%%%%%%%%%%%%%%%%%%%%%%%%%%%%%%%%%%%%%%%%%%

\section{A General Renewal Process}\label{sec:general_result}
\begin{figure}[t]
\centering
\includegraphics[width=0.5\textwidth]{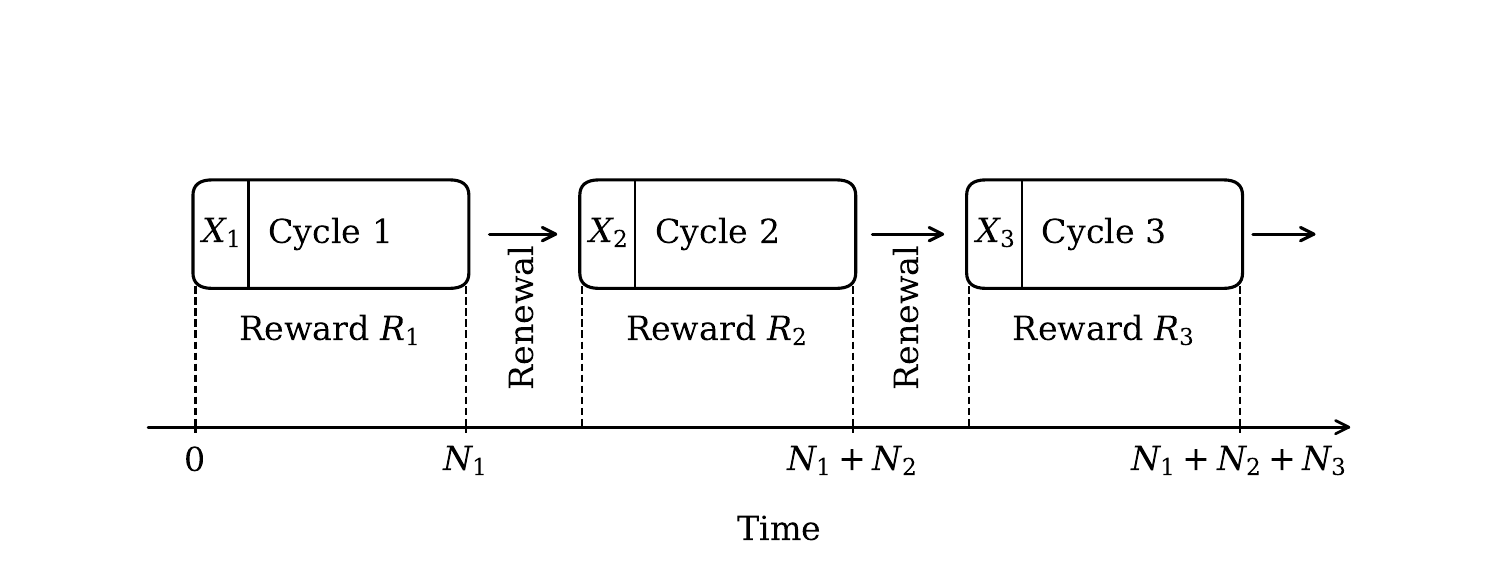}
\caption{Schematic of the multi-phase alternating renewal process. The system evolves 
through a sequence of i.i.d. cycles. At the start of each cycle $i$, a sign variable $X_i \in \{-1, +1\}$ 
is determined, defining the phase. Each cycle is characterized by a discrete duration $N_i$ 
(number of time steps) and a total reward $R_i$. The boxes represent the black-box internal 
dynamics. The phase $X_i$ is reset and redrawn at every renewal point.}
\label{fig:Schematic}
\end{figure}

We consider a stochastic process evolving in independent and identically distributed (i.i.d.) 
cycles indexed by $i=1,2,\dots$. Each cycle is characterized by a random duration $N_i>0$, a 
total reward (current magnitude) $R_i$, and an auxiliary sign variable $X_i \in \{-1,+1\}$ representing directional bias.
The cycle duration $N_i$ may be taken as either a discrete number of steps
or a continuous elapsed time; the renewal-reward identities used below hold
in either case, and discreteness of $N_i$ is nowhere assumed in this section.
The ``discrete-time'' qualifier attached to the Proesmans--Van den Broeck
inequality below refers only to the intrinsically two-valued variable $X_i$,
not to $N_i$.
We make no assumptions regarding the internal dynamics or the phase structure of the reward 
production; $R_i$ may arise from any functional of the internal trajectory (see Fig. \ref{fig:Schematic}). 
The pairs $(N_i, R_i)$ are i.i.d. across cycles with finite first and second moments, and the auxiliary 
variables $X_i$ are i.i.d. and independent of $(N_i, R_i)$, with:
\begin{equation}
\mathbb{E}[X_i] = m_X, \quad \operatorname{Var}(X_i) = 1 - m_X^2.
\end{equation}
The current increment for cycle $i$ factorizes as $\Delta J_i = X_i R_i$. Let $J(t) = \sum_{i=1}^{\nu(t)} \Delta J_i$ 
be the integrated current up to time $t$, where $\nu(t)$ is the number of completed cycles. 
Stationary renewal–reward theory yields the mean current $\bar{j}$ and asymptotic variance $\sigma_j^2$ \cite{Smith1958}:
\begin{eqnarray}
\bar{j} &=& \lim_{t\to\infty}\frac{\mathbb{E}[J(t)]}{t} = \frac{\mathbb{E}[X] \mathbb{E}[R]}{\mathbb{E}[N]}, \\
\sigma_j^2 &=& \lim_{t\to\infty}
\frac{\operatorname{Var}(J(t))}{t} = \frac{\operatorname{Var}(\Delta J - \bar{j}N)}{\mathbb{E}[N]}.
\end{eqnarray}
To evaluate the variance of the drifted reward, $\operatorname{Var}(XR - \bar{j}N)$, we note that 
since $\mathbb{E}[XR - \bar{j}N] = 0$, the variance equals the second moment. 
Defining $\bar{r} = \mathbb{E}[R]/\mathbb{E}[N]$, we obtain the decomposition (see Appendix~\ref{app:variance_decomposition}):
\begin{equation}
\operatorname{Var}(\Delta J - \bar{j}N) = \operatorname{Var}(X) \mathbb{E}[R^2] + m_X^2 \operatorname{Var}(R - \bar{r}N).
\end{equation}
Here and in what follows, $(X,R,N)$ denotes a generic i.i.d.\ copy of the
cycle variables $(X_i,R_i,N_i)$, and $\Delta J = XR$ is the corresponding
representative per-cycle current increment; the cycle index $i$ is dropped
whenever a generic (representative) cycle, rather than a specific one in
the sequence, is meant. This identity demonstrates that the variance
consists of a term driven by the auxiliary sign fluctuations and a term
driven by the internal variability of the reward-duration ratio; note that
$N$ and $R$ are not assumed independent, and it is precisely their
correlation that this second term captures --- independence is assumed
only for $X$ relative to the pair $(N,R)$.
Since the second term is non-negative, the current variance is bounded 
below by the statistics of the auxiliary process:
\begin{equation}
\sigma_j^2 \ge \frac{\operatorname{Var}(X)}{\mathbb{E}[N]} \mathbb{E}[R^2].
\end{equation}
This lower bound is model-independent, saturating when the reward magnitude 
is perfectly correlated with the cycle duration ($R = \bar{r}N$). 

We consider auxiliary sequences $\{X_i\}$ that are i.i.d.\ and
$\{-1,+1\}$-valued, and we write $p\equiv\mathbb{P}(X=+1)=(1+m_X)/2\in(0,1)$. Such a
sequence is the step sequence of a biased random walk, i.e.\ of a stationary,
time-homogeneous Markov chain whose one-step current $X_i$ is odd under time
reversal, and its dimensionless entropy production per cycle is
$\bar s_X=(2p-1)\ln[p/(1-p)]$ (its meaning and the convention adopted are
discussed below Eq.~\eqref{eq:linear_tur_final}). For such sequences the
discrete-time Proesmans--Van den Broeck (PV) inequality \cite{Proesmans2017}
holds, $m_X^2/\operatorname{Var}(X)\le\frac12(e^{\bar s_X}-1)$; for
completeness, a direct proof is given in Appendix~\ref{app:PV_binary}.

Combining this with the variance bound yields the modified thermodynamic uncertainty relation:
\begin{equation}\label{eq:linear_tur_final}
\frac{\bar{j}^2}{\sigma_j^2} \le \frac{1}{2}(e^{\bar{s}_X} - 1) 
\frac{\mathbb{E}[R]^2}{\mathbb{E}[N] \mathbb{E}[R^2]}.
\end{equation}

Two points concerning Eq.~\eqref{eq:linear_tur_final} deserve clarification.
First, the discrete-time PV inequality invoked above is,
mathematically, an unconditional relation between the mean and variance of
\emph{any} $\{-1,+1\}$-valued random variable and the quantity
$\bar{s}_X\equiv(2p-1)\ln[p/(1-p)]$, which is simply the Kullback--Leibler (KL) divergence
between the Bernoulli$(p)$ law of $X$ and its time-reversed counterpart
Bernoulli$(1-p)$; as such it holds regardless of any physical interpretation
and requires only the observed bias $p$. Second, the identification of
$\bar{s}_X$ with a genuine entropy production follows from the standard
local-detailed-balance assumption of stochastic thermodynamics applied to
the reorientation (tumble) event itself: if the ratio $p/(1-p)$ is set by
the ratio of forward-to-backward rates 
of the microscopic mechanism selecting the new run direction, with the
direction $X$ odd under time reversal, then $\bar{s}_X$ is the stochastic
entropy produced (in units of $k_B$) by that degree of freedom,
$\bar{s}_X = \ln[P(\text{path})/P(\text{reversed path})]$ restricted to the
marginal $X$-sequence. 
This can be made fully explicit. For a trajectory of $n$ i.i.d.\ realizations
$X_1,\dots,X_n$ of the auxiliary variable, each with $\mathbb{P}(X_i=+1)=p$,
the probability of the forward path relative to its time-reversed
counterpart (in which every $X_i\to-X_i$) is
\[
\frac{P(X_1,\dots,X_n)}{P(-X_1,\dots,-X_n)}
= \prod_{i=1}^n \left(\frac{p}{1-p}\right)^{X_i},
\]
using $\mathbb{P}(X_i=x)=p^{(1+x)/2}(1-p)^{(1-x)/2}$ for $x\in\{-1,+1\}$. The
stochastic entropy production along the trajectory is therefore
\[
\Delta s_X = \ln\frac{P(X_1,\dots,X_n)}{P(-X_1,\dots,-X_n)}
= \sum_{i=1}^n X_i\,\ln\frac{p}{1-p},
\]
and its mean per cycle follows directly from $\mathbb{E}[X_i]=2p-1$:
\[
\bar{s}_X = \mathbb{E}[\Delta s_X]/n = (2p-1)\ln\frac{p}{1-p},
\]
confirming that, with $X$ odd under time reversal, $\bar s_X$ is the average
stochastic entropy production of the directional sequence, in the sense
of~\cite{Seifert2012}.

A remark on conventions is in order. Here time reversal acts on the
directional variable by a sign flip, $X\to-X$: $X$ is taken to be odd, like the
displacement $\Delta J=XR$ it generates, and $\bar s_X$ is the entropy
production of the auxiliary biased walk with steps $X_i$, which is the quantity
entering the PV inequality. This differs from the convention often adopted for
run-and-tumble particles, in which the orientation is even under time reversal
\cite{Garcia-MillanPruessner2021}; in that convention the i.i.d.\ tumble
sequence by itself carries zero entropy production, since the orientation chain
satisfies detailed balance with respect to its stationary distribution, and any
dissipation is associated with other degrees of freedom, such as the
self-propulsion. The two conventions thus assign different entropy productions
to the same physical process, and the PV inequality for $X$ is applied in the
odd one. Equation~\eqref{eq:linear_tur_final} is therefore a bound in terms of
$\bar s_X$ as defined here, irrespective of any further physical
interpretation. If, in addition, the time reversal of the full underlying
process also flips $X$, so that coarse-graining to the $X$-sequence commutes
with time reversal, the coarse-graining (data-processing) inequality
\cite{KawaguchiNakayama2013,BoCelani2017} implies that $\bar s_X$ is a lower
bound on the total entropy production of the full process, and
Eq.~\eqref{eq:linear_tur_final} remains a valid, if conservative, relation even
when the full microscopic dissipation is inaccessible or left unspecified. When
this condition is not met (e.g.\ for an even orientation), $\bar s_X$ should be
read as the entropy production of the directional choice in the odd convention,
i.e.\ as a measure of the asymmetry between $p$ and $1-p$, and no inequality
with the total physical entropy production is implied.

The independence of $X$ and $R$ also fixes the distributional structure of
$\Delta J = XR$: if $R$ has density $f_R(r)$, then
\[
f_{\Delta J}(j) = p\, f_R(j) + (1-p)\, f_R(-j),
\]
so that $\Delta J$ is a weighted two-sided version of the reward distribution. 
When $R$ is non-negative, its positive and negative tails are exact scaled 
copies of one another in ratio $p/(1-p)$; for general (sign-indefinite) $R$ — as 
in our AR(1) example of Sec.~\ref{sec:example2} — this simple tail relation 
need not hold, though the density formula above still applies without modification. 
For non-negative $R$, this provides a simple, testable signature of the 
independence assumption in experimental or simulated current-increment data.
In particular, a single-cycle increment $\Delta J$ with a Gaussian law of
non-zero mean cannot be written as $XR$ with $X$ independent of $R$ for any
bias $p\in(0,1)$. Indeed, denoting by $g_s$ and $g_a$ the even and odd parts of
$f_{\Delta J}$, the formula above gives $f_R=g_s+g_a/(2p-1)$, whose
non-negativity requires $|g_a/g_s|\le|2p-1|<1$, whereas for a Gaussian
$\mathcal N(\mu,\sigma^2)$ one has $g_a/g_s=\tanh(\mu j/\sigma^2)$, which exceeds
this value at large $|j|$ if $\mu\neq0$. A zero-mean Gaussian increment, by
contrast, is compatible with the structure (take $R$ symmetric Gaussian and any
$p$). This restriction concerns the single-cycle increment only; the integrated
current $J(t)$ is asymptotically Gaussian with non-zero mean, as 
usual.

The TUR relation \eqref{eq:linear_tur_final} applies to any square-integrable renewal process 
with factorized increments. Unlike the derivation in \cite{ShreshthaHarris2019}, which assumes 
a specific microscopic scaling where reward variance grows linearly with run duration, the present 
formulation relies only on the renewal structure. Consequently, the bound derived here is valid for 
arbitrary internal dynamics, including non-additive or highly correlated reward processes, provided 
the second moments remain finite. 
%%%%%%%%%%%%%%%%%%%%%%%%%%%%%%%%%%%%%%%%%%%%%%%%%%%%%%%%%%%%%%
\section{Small-Bias Expansion: Linear-Response Theory} \label{sec:small_bias}
We now analyze the TUR relation \eqref{eq:linear_tur_final} in the linear-response regime, corresponding 
to the limit of a weak directional bias \cite{Barato2015,DechantSasa2020}. Since the internal micro-mechanics 
of the underlying system are treated as a black-box, we quantify the weak thermodynamic driving force directly 
through the macroscopic asymmetry of the cycles. Specifically, we parameterize the probability of the auxiliary 
directional variable taking a forward orientation ($X = +1$) as $p = 1/2 + \varepsilon$, with $|\varepsilon| \ll 1$. 
In this linear-response regime the auxiliary process characteristics satisfy:
\begin{equation}
\mathbb{E}[X] = 2\varepsilon, \quad \operatorname{Var}(X) = 1 - 4\varepsilon^2.
\end{equation}
The dimensionless entropy production per cycle associated with the auxiliary process is now given by
\begin{equation}
\bar{s}_X = (2p-1)\ln\frac{p}{1-p} = 8\varepsilon^2 + \mathcal{O}(\varepsilon^4).
\end{equation}
In this regime, the entropic term in the bound becomes $e^{\bar{s}_X} - 1 = 8\varepsilon^2 + \mathcal{O}(\varepsilon^4)$. 
Substituting this expansion into the generalized TUR derived in the previous section yields
\begin{equation}
\frac{\bar{j}^2}{\sigma_j^2} \le 4\varepsilon^2 \frac{\mathbb{E}[R]^2}{\mathbb{E}[N] \left( \operatorname{Var}(R) + \mathbb{E}[R]^2 \right)} + \mathcal{O}(\varepsilon^4).
\end{equation}
Using $\mathbb{E}[X] = 2\varepsilon$ the above equation can be written in the informative form:
\begin{equation} 
\frac{\bar{j}^2}{\sigma_j^2} \le \frac{\mathbb{E}[X]^2}{\mathbb{E}[N]} \cdot \frac{1}{1 + \operatorname{Var}(R)/\mathbb{E}[R]^2} + \mathcal{O}(\mathbb{E}[X]^4).
\end{equation}
This equation shows that the current precision scales quadratically with the bias, as expected from 
linear-response theory. Crucially, the renewal statistics enter through a dimensionless suppression 
factor, $1 / (1 + \operatorname{Var}(R)/\mathbb{E}[R]^2)$, which reduces the achievable precision 
whenever cycle-length or reward fluctuations are present. 

In the deterministic-cycle limit where internal reward fluctuations vanish ($\operatorname{Var}(R) \to 0$), the 
bound reduces to
\begin{equation}
\frac{\bar{j}^2}{\sigma_j^2} \le \frac{\mathbb{E}[X]^2}{\mathbb{E}[N]} + \mathcal{O}(\mathbb{E}[X]^4),
\end{equation}
recovering the standard linear-response TUR, ${\bar{j}^2}/{\sigma_j^2} \le {\Sigma}/{2}$, where 
$\Sigma = \bar{s}_X / \mathbb{E}[N]$ represents the mean entropy production rate of the auxiliary 
process. This confirms that our generalized renewal bound is a consistent extension of universal 
thermodynamic constraints to processes with non-trivial temporal structures.
%%%%%%%%%%%%%%%%%%%%%%%%%%%%%%%%%%%%%%%%%%%%%%%%%%%%%%%%%%%%%%
\section{ A Two-Phase Run--Rest Renewal Process}\label{sec:example1}
Motivated by classical velocity-jump models incorporating resting states \cite{TaylorKing2015}, we 
consider a renewal model consisting of two independent phases within each cycle. In the first phase, the 
particle undergoes a biased random walk (run phase) and accumulates displacement, while in the second 
phase, it enters a waiting (rest) state in which no motion and hence no current accumulation occurs. Each 
full run--wait cycle is punctuated by an instantaneous tumble that resets the direction of motion.

This framework shares structural similarities with the generalized velocity-jump processes investigated 
in \cite{TaylorKing2015}, which distinguish between active phases contributing to displacement and passive 
phases influencing renewal time. While \cite{TaylorKing2015} utilizes Cattaneo approximations to demonstrate 
how dwell-time variances govern dispersion, our framework addresses the fundamental thermodynamic 
uncertainty. Specifically, we show how phase-specific statistics explicitly constrain the current precision via 
a modified thermodynamic uncertainty relation.
%%%%%%%%%%%%%%%%%%%%%%%%%%%%%%%%%%%%%%

We consider a discrete-time process in one spatial dimension where each cycle $i$ contains two consecutive
phases. At the start of each cycle (the tumble) a heading $X\in\{+1,-1\}$ is drawn, with
$\mathbb{P}(X=+1)=p$ and $\mathbb{P}(X=-1)=1-p$, independently of all other random variables.
In Phase 1 (active), the phase terminates with probability $r_1$ at each step; the terminating step carries
no displacement. Otherwise, with probability $1-r_1$, the particle performs a displacement step $X\xi_k$, where
the increments $\xi_k$ are i.i.d., independent of $X$, with
\begin{equation}
\xi_k =
\begin{cases}
+1 & \text{with probability } p,\\
-1 & \text{with probability } 1-p.
\end{cases}
\end{equation}
In Phase 2 (inactive), no displacement occurs, and the phase terminates with probability $r_2$ at each step.
The phase durations (each counting its terminating step) are thus geometric random variables:
\begin{equation}
\mathbb{P}(N_1=n)=(1-r_1)^{n-1}r_1, \; \mathbb{P}(N_2=n)=(1-r_2)^{n-1}r_2,
\end{equation}
yielding the mean durations $\mathbb{E}[N_1]=1/r_1$ and $\mathbb{E}[N_2]=1/r_2$. The total cycle
duration $N = N_1 + N_2$ has the mean $\mathbb{E}[N] = 1/r_1 + 1/r_2$. The number of displacement steps
in a run is $K=N_1-1$.

%%%%%%%%%%%%%%%%%%%%%%%%%%%%%%%%%%%%%%
To map this onto our general renewal-reward structure, the auxiliary sign 
variable $X$ is the heading drawn at the start of the cycle. The entropy 
production associated with this choice is $\bar{s}_X = (2p-1)\ln(p/(1-p))$. 
Here, as in assumption A3, $\bar s_X$ is defined with the heading odd under time
reversal, $X\to-X$. Only the displacement steps of the active phase contribute to the reward,
$R = \sum_{k=1}^{N_1-1} \xi_k$, so that $\Delta J = XR$ with $X$ independent of $(N,R)$, as required.
Using the laws of total expectation and variance, we find:
\begin{equation}
\begin{aligned}
\mathbb{E}[R]&=\frac{(2p-1)(1-r_1)}{r_1},\\
\operatorname{Var}(R)&=(1-r_1)\left[\frac{4p(1-p)}{r_1}+\frac{(2p-1)^2}{r_1^2}\right].
\end{aligned}
\end{equation}
The raw second moment of the reward is thus:
\begin{equation}
\begin{aligned}
\mathbb{E}[R^2]&=\operatorname{Var}(R)+\mathbb{E}[R]^2\\
&=(1-r_1)\left[\frac{4p(1-p)}{r_1}+(2p-1)^2\frac{2-r_1}{r_1^2}\right].
\end{aligned}
\end{equation}
Substituting these into our generalized bound, we obtain the explicit precision constraint:
\begin{equation} \label{eq:run_rest_bound}
\frac{\bar{j}^2}{\sigma_j^2} \le \frac{(e^{\bar{s}_X}-1)}{2} \frac{(2p-1)^2 (1-r_1)\, r_1 r_2}{(r_1+r_2)\left[4p(1-p)r_1+(2p-1)^2(2-r_1)\right]}.
\end{equation}

%%%%%%%%%%%%%%%%%%%%%%%%%%%%%%%%%%%%%%
\begin{figure}[t]
\centering
\includegraphics[width=0.5\textwidth]{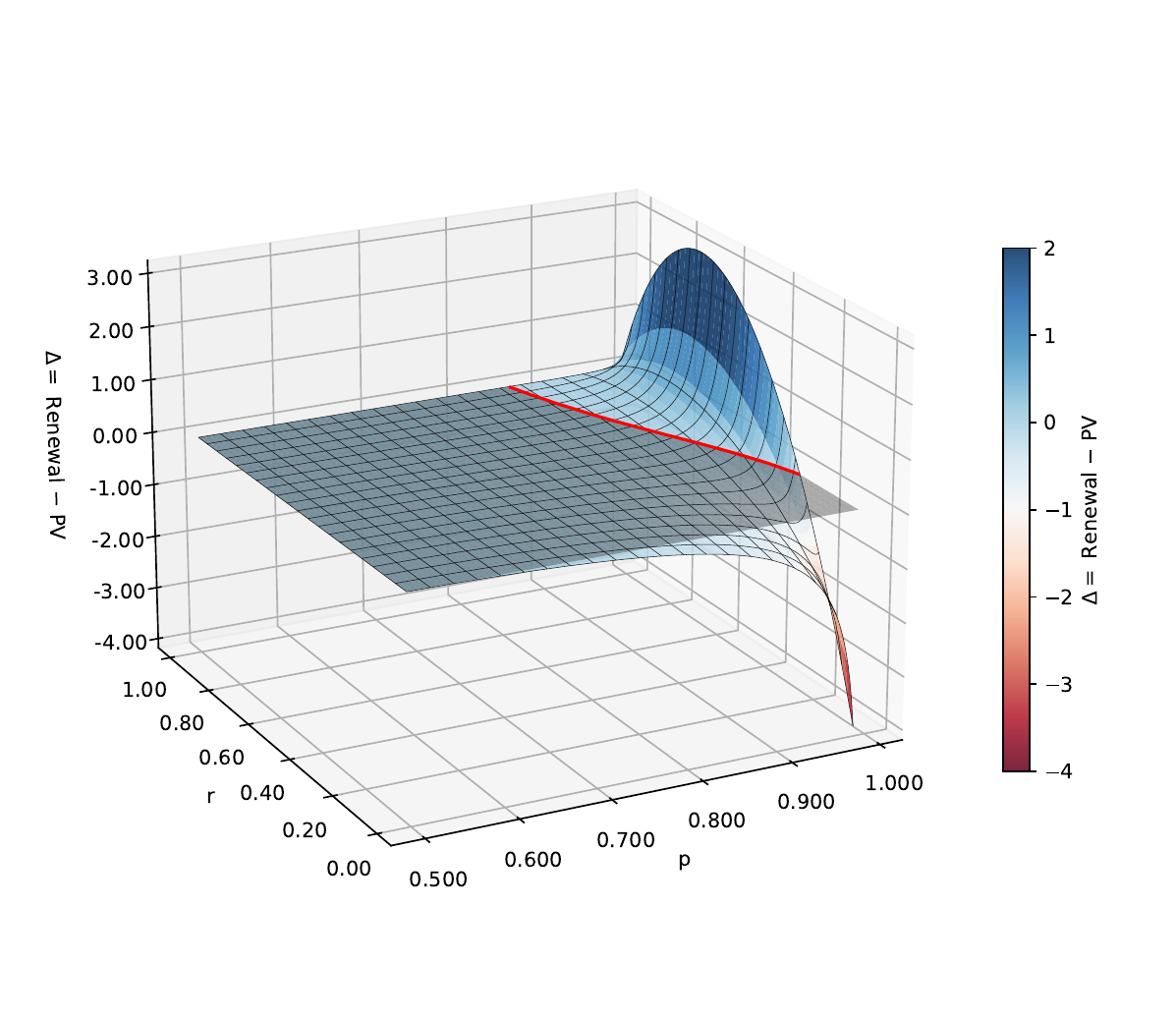}
\caption{Comparison of the generalized renewal bound and the
Proesmans--Van den Broeck (PV) bound for the two-phase run--rest
process with $r_1=r_2=r$. The surface shows the difference
$\Delta\equiv(\text{Renewal Bound})-(\text{PV Bound})$ for $p\ge1/2$,
with the PV bound evaluated in the extended state space (heading even
under time reversal) using
$\bar s_{\rm tot}=\frac{1-r}{2}(2p-1)\ln\frac{p}{1-p}$. 
The transparent plane at $\Delta=0$ provides a reference for the sign
of $\Delta$; the red curve marks the intersection of the surface with
this plane, i.e.\ the crossover $r_c(p)$ between the two bounds: red
regions ($\Delta<0$) are where the renewal bound is tighter, blue
regions ($\Delta>0$) where the PV bound is tighter.
The renewal bound is tighter at long cycles (small $r$) for every $p$,
since the renewal/PV ratio tends to zero as $r\to0$. Both bounds vanish
as $r\to1$; their ratio there,
$(2p-1)^2(e^{\sigma_0}-1)/\sigma_0$, crosses unity at
$p^*\simeq0.854$. Consequently, for $p<p^*$ the renewal bound is
tighter at all $r$, while for $p>p^*$ the PV bound is tighter above a
crossover $r_c(p)$ that decreases as $p$ increases ($r_c\approx0.82$ at
$p=0.88$, $\approx0.24$ at $p=0.99$). The figure is symmetric under
$p\to1-p$, reflecting the invariance of both $(2p-1)^2$ and $\bar s_X$.
The colour scale is clipped at $\Delta=-4$ and $\Delta=2$; the true
magnitude at extreme bias exceeds the plotted range.}
\label{fig:bound_comparison}
\end{figure}
%%%%%%%%%%%%%%%%%%%%%%%%%%%%%%%%%%%%%%

To provide a benchmark for this result, we compare it with the PV bound evaluated with
the total mean entropy production rate of the process, computed as in
Ref.~\cite{ShreshthaHarris2019} from the ratio of the probabilities of a trajectory and
of its time reversal in the extended state space (position and heading), with the heading
even under time reversal. In this convention the tumbles contribute nothing on average,
since the heading changes from positive to negative as often as from negative to positive,
and the entropy production comes entirely from the displacing steps, each of which
produces on average $\sigma_0\equiv(2p-1)\ln[p/(1-p)]$. The fraction of time steps that
are displacing is $(1-r_1)r_2/(r_1+r_2)$, so
$\bar s_{\rm tot}=\frac{(1-r_1)r_2}{r_1+r_2}\,\sigma_0$. For the special case $r_1=r_2=r$
this gives $\bar s_{\rm tot}=\frac{1-r}{2}(2p-1)\ln\frac{p}{1-p}$ and
\begin{equation} \label{eq:naive_pv}
\frac{\bar{j}^2}{\sigma_j^2} \le \frac{1}{2} \left[
\left(\frac{p}{1-p}\right)^{\frac{1-r}{2}(2p-1)} - 1 \right].
\end{equation}
Note that the heading is treated as even here, for the total entropy production of the
process, whereas $\bar s_X$ in Eq.~\eqref{eq:run_rest_bound} is defined with the heading
odd (assumption A3); the two quantities are different and no
inequality between them is implied.\footnote{If the 
heading is instead treated as odd under time reversal, one obtains the
$p$-dependent value $\bar s_{\rm tot}=\frac r2\sigma_0$, for which the PV inequality is
not guaranteed to hold \cite{ShreshthaHarris2019}.}
%%%%%%%%%%%%%%%%%%%%%%%%%%%%%%%%%%%%%%

Comparing Eq.~\eqref{eq:run_rest_bound} with Eq.~\eqref{eq:naive_pv} across the
$(p,r)$ parameter space (see Fig.~\ref{fig:bound_comparison}), we find that the
two bounds are complementary only at large directional bias. Writing
$\sigma_0\equiv\bar s_X=(2p-1)\ln[p/(1-p)]$, the ratio of the renewal bound to the
PV bound tends to zero as $r\to0$, because the renewal bound vanishes linearly in
$r$ while the PV bound remains finite, and tends to
$(2p-1)^2(e^{\sigma_0}-1)/\sigma_0$ as $r\to1$, where both bounds vanish as $(1-r)$.
This limiting ratio is smaller than unity for $p<p^*\simeq0.854$ and larger for
$p>p^*$. Consequently, for moderate bias ($p<p^*$) the renewal bound is tighter
($\Delta<0$) at all $r$, as seen in the red regions of
Fig.~\ref{fig:bound_comparison}. For larger bias ($p>p^*$) the renewal bound is
tighter at long cycles, while the PV bound is tighter above a crossover $r_c(p)$,
i.e.\ at short cycles, with $r_c$ decreasing as $p$ increases ($r_c\approx0.82$ at
$p=0.88$ and $\approx0.24$ at $p=0.99$). The figure is symmetric under $p\to1-p$,
reflecting the invariance of both $(2p-1)^2$ and $\bar s_X$ under this exchange.

The two bounds use different statistical inputs: the PV bound depends only on the
mean entropy production rate, whereas the renewal bound retains the cycle-level
statistics of the reward through the prefactor
$\mathbb{E}[R]^2/(\mathbb{E}[N]\,\mathbb{E}[R^2])$. This accounts for the two
limits. As $r\to0$ the cycles become long and the current precision vanishes
linearly in $r$; the renewal bound has the same linear scaling, through the factor
$1/\mathbb{E}[N]=r/2$, whereas the PV bound tends to a finite value, since the
mean entropy production rate per time step remains finite in this limit. As
$r\to1$ the mean number of displacing steps per cycle,
$\mathbb{E}[K]=(1-r)/r$, vanishes and the current precision vanishes with it,
again linearly in $1-r$. Both bounds reproduce this: the renewal bound through the
factor $(1-r)$ in $\mathbb{E}[R]$, and the PV bound through the fraction of
displacing time steps, $(1-r)/2$, entering $\bar s_{\rm tot}$. They differ only in
the coefficient of this linear vanishing, which is what produces the crossover for
$p>p^*$. We also note that the duration of the rest phase enters
Eq.~\eqref{eq:run_rest_bound} only through $\mathbb{E}[N]$, since the term of
the variance decomposition that carries $\operatorname{Var}(N)$ and
$\operatorname{Cov}(R,N)$ has been discarded; the bound is therefore
insensitive to the variance of the rest time.
%%%%%%%%%%%%%%%%%%%%%%%%%%%%%%%%%%%%%%

In this section we have formulated the two-phase run--rest model in one
spatial dimension, with the particle displaced by $X\xi_k$ at each displacing 
step of the active phase, where $X$ is the heading drawn at the tumble 
and $\xi_k\in\{-1,+1\}$. The generalization to higher dimensions 
admits two distinct interpretations, and it is important to distinguish them. 
%%%%%%%%%%%%%%%%%%%%%%%%%%%%%%%%%%%%%%

First, one may consider a particle that moves along several independent axes 
simultaneously — for instance, a particle on a $d$-dimensional grid that, at each 
tumble, independently decides its bias along each axis. All directional decisions occur 
together at each tumble, while a single shared run duration and displacement magnitude 
apply identically to every axis; only the sign of the resulting motion differs axis by axis. 
In this case the auxiliary variable is promoted to a vector, while the reward retains its scalar character. 
The resulting variance decomposition remains exact, with $\operatorname{Var}(X)$ generalizing to 
the covariance matrix of the directional biases plus an additional term reflecting the correlation the 
shared run statistics induce between axes; because the directional biases themselves remain 
mutually independent, the associated entropy production is additive across axes, and the 
generalization closes into an explicit bound.
%%%%%%%%%%%%%%%%%%%%%%%%%%%%%%%%%%%%%%

Second, and more realistically, one may consider a particle that, at each tumble, selects a single 
heading among $d$ possible directions — as in a swimming or crawling organism reorienting itself. 
Here there is only one directional decision per tumble, but it has more than two possible outcomes, 
and the different headings are intrinsically coupled: choosing a strong bias toward one heading 
necessarily reduces the probability of the others, since the probabilities must sum to unity. This 
coupling does not affect the variance decomposition itself, which remains valid for an arbitrary 
covariance structure; rather, it invalidates the axis-by-axis argument used to establish the entropic 
bound in the independent case. Closing this gap requires a genuinely multivariate discrete-time 
thermodynamic uncertainty relation, valid for correlated directional components, and remains an open problem.
%%%%%%%%%%%%%%%%%%%%%%%%%%%%%%%%%%%%%%

We restrict to one dimension here for clarity and analytical tractability, as our primary aim is to illustrate the 
renewal structure rather than the specifics of the spatial geometry.
%%%%%%%%%%%%%%%%%%%%%%%%%%%%%%%%%%%%%%%%%%%%%%%%%%%%%%%%%%%
\section{Non-Markovian Run Dynamics: Correlated Internal Reward Process}\label{sec:example2}
To illustrate the generality of the renewal framework, we consider a two-phase run--rest process where the 
internal dynamics during the run phase exhibit temporal correlations \cite{Pigolotti2017}. This example 
violates the IID increment assumption inherent in many previous studies \cite{ShreshthaHarris2019} while 
remaining strictly within the scope of our renewal formulation.

The process consists of alternating rest (tumble) and run phases. Each renewal cycle $i$ contains a rest 
event of fixed duration (one time step) and a run event of random duration $L_i \in \mathbb{N}$, such that 
the cycle duration is $N_i = 1 + L_i$. The run lengths $L_i$ are i.i.d. random variables with finite first and 
second moments. Directional bias is encoded by a binary variable $X \in \{-1, +1\}$ with $\mathbb{E}[X] = 2p-1$, which 
determines the sign of the current accumulated during the run. As in the previous examples, the auxiliary 
entropy $\bar{s}_X$ represents the thermodynamic cost associated with establishing the directional bias $p$ at each renewal event.

During a run of length $L$, the accumulated reward is $R=R_{\text{run}} = \sum_{i=1}^{L} r_i$. We assume 
the increments $\{r_i\}$ follow a stationary autoregressive process of order one, AR(1):
\begin{equation}
r_i = \mu + \rho (r_{i-1}-\mu) + \xi_i, \qquad |\rho|<1,
\end{equation}
where $\{\xi_i\}$ are independent Gaussian innovations with $\text{Var}(\xi_i) = \sigma_\xi^2$. In the stationary 
regime, the increments have mean $\mu$ and a covariance structure $\text{Cov}(r_i, r_j) = \sigma_r^2 \rho^{|i-j|}$, 
where $\sigma_r^2 = \sigma_\xi^2 / (1-\rho^2)$. The parameter $\rho$ controls the strength of temporal correlations. 
For $\rho=0$, the increments become IID, recovering the standard diffusive case. For $\rho \neq 0$, the increments 
are correlated, and the variance of the accumulated reward is no longer linear in the run length.

The mean reward for a run of length $L$ is $\mathbb{E}[R_{\text{run}}|L] = \mu L$. Using the AR(1) covariance 
structure, the conditional variance is given by:
\begin{eqnarray}
\text{Var}(R_{\text{run}}|L) &=& \sigma_r^2 \left[ L + 2\sum_{k=1}^{L-1}(L-k)\rho^k \right] \nonumber \\
&=& \sigma_r^2 \left[ \frac{L(1+\rho)}{1-\rho} - \frac{2\rho(1-\rho^L)}{(1-\rho)^2} \right].
\end{eqnarray}
By the law of total variance, the total variance of the run reward is:
\begin{equation}
\text{Var}(R_{\text{run}}) = \mathbb{E}[\text{Var}(R_{\text{run}}|L)] + \mu^2 \text{Var}(L).
\end{equation}
Defining the total cycle reward as $\Delta J = X R_{\text{run}}$, the mean and variance follow from the factorization 
properties discussed in 
Section \ref{sec:general_result}:
\begin{eqnarray}
\mathbb{E}[\Delta J] &=& (2p-1)\mu \mathbb{E}[L], \\
\text{Var}(\Delta J) &=& \text{Var}(R_{\text{run}}) + 4p(1-p)(\mu \mathbb{E}[L])^2.
\end{eqnarray}

To obtain a fully explicit result, we consider geometrically distributed run 
lengths, $\mathbb{P}(L=\ell) = r(1-r)^{\ell-1}$ for $\ell \ge 1$. Utilizing the memoryless property of the 
geometric distribution, the expected conditional variance simplifies to:
\begin{equation}
\mathbb{E}\left[ L + 2\sum_{k=1}^{L-1}(L-k)\rho^k \right] = \frac{1}{r} + \frac{2(1-r)\rho}{r[1-(1-r)\rho]}.
\end{equation}
Substituting this into our generalized bound yields a closed-form analytical expression for the precision 
constraint as a function of the correlation strength $\rho$, the bias $p$, and the run-length statistics $r$:
\begin{equation} \label{eq:ar1_bound}
\frac{\bar{j}^2}{\sigma_j^2} \le \frac{1}{2}(e^{\bar{s}_X}-1)
\frac{\mu^2 \mathbb{E}[L]^2}{(1+\mathbb{E}[L])\left[ \text{Var}(R_{\text{run}}) + \mu^2 \mathbb{E}[L]^2 \right]}.
\end{equation}

For the case of geometrically distributed run lengths with parameter $r$, the explicit variance of the run 
reward is obtained by combining the expected internal AR(1) fluctuations with the variance of the run duration, yielding:
\begin{equation}
\operatorname{Var}(R_{\text{run}}) = \frac{\sigma_r^2}{r} \left( 1 + \frac{2\rho(1-r)}{1-\rho(1-r)} \right) + \mu^2 \frac{1-r}{r^2}.
\end{equation}
This expression, together with $\mathbb{E}[L]=1/r$, allows for a fully analytical evaluation of the precision 
bound in Eq.~\eqref{eq:ar1_bound}.

The strength of this example lies in its departure from the standard IID assumption. By incorporating an AR(1) 
structure, we demonstrate that the generalized renewal bound is robust even for non-Markovian run 
dynamics \cite{HasegawaVu2019} where persistence ($\rho > 0$) or anti-persistence ($\rho < 0$) dictates 
current accumulation. This "stress test" illustrates that microscopic scaling laws—such as the linear variance growth 
often assumed in earlier literature—are not required for the validity of the bound. Instead, the renewal framework 
successfully reduces complex, correlated internal dynamics into cycle-level moments, providing a robust tool for 
characterizing active matter systems where memory effects and colored noise are physically prevalent.
%%%%%%%%%%%%%%%%%%%%%%%%%%%%%%%%%%%%%%%%%%%%%%%%%%%%%%%%%%%
\section{Summary and Discussion}
In this work, we have addressed the fundamental challenge of bounding current fluctuations in non-equilibrium 
systems that do not satisfy the standard assumptions of time-homogeneous Markovian dynamics. While the TUR 
has provided a transformative framework for understanding the trade-off between precision and dissipation, its 
application to processes with complex internal dynamics—such as those found in active matter—has often 
required specific microscopic assumptions that limit its universality. Building upon the renewal-reward framework 
initially applied to run-and-tumble motion by Shreshtha and Harris \cite{ShreshthaHarris2019}, we have derived a 
generalized bound that decouples current precision from the specific internal scaling of renewal cycles.

By utilizing a rigorous variance decomposition, we have demonstrated that the precision of a renewal-reward process 
is constrained by a combination of the entropy production of the directional choice and a model-independent prefactor. 
This prefactor isolates the fundamental current fluctuations from the cycle-level correlations between duration and 
displacement. Our analysis in the linear-response regime reveals that this structure identifies a 
universal ``renewal penalty''--a suppression of precision arising from fluctuations of the cycle reward,
whether intrinsic or induced by fluctuating run durations. 

The robustness of this formulation was demonstrated through two distinct
examples. First, our analysis of the two-phase run--rest model provides a
precision-based complement to classical velocity-jump dispersion theories
\cite{TaylorKing2015}. At long cycles the current precision vanishes
linearly in the rate $r$ at which phases terminate, and the renewal bound
tracks this vanishing through the cycle statistics of the reward, whereas
the PV bound, which depends only on the mean entropy production rate,
remains finite and therefore loose.

The run--rest model of Sec.~\ref{sec:example1} provides a concrete
illustration of why standard TURs can be loose for renewal processes. As
shown in Fig.~\ref{fig:bound_comparison}, for long cycles ($r\ll1$) the 
renewal bound is tighter than the PV bound evaluated with the total
entropy production (heading even) for every $p$. The reason is
that the PV bound depends only on the mean entropy production rate,
$\bar s_{\rm tot}=(1-r)\sigma_0/2\to\sigma_0/2$ as $r\to0$, which remains
finite, whereas the true current precision vanishes linearly in $r$. The
renewal bound carries the same linear vanishing through the cycle-level
statistics of the reward: at small $r$ the mean number of displacing
steps per cycle, $(1-r)/r$, shrinks while the mean cycle duration,
$2/r$, grows, and both effects suppress the prefactor
$\mathbb E[R]^2/(\mathbb E[N]\mathbb E[R^2])$ so that the bound tracks
the true precision. At short cycles both bounds vanish linearly in
$1-r$, and their ratio approaches
$(2p-1)^2(e^{\sigma_0}-1)/\sigma_0$, so for $p>p^*$ the PV bound becomes
tighter above a crossover $r_c(p)$ that decreases with $p$.

Second, by incorporating AR(1) internal correlations, we have proven that the 
renewal-reward formulation remains valid even when the underlying dynamics 
are non-Markovian and the i.i.d. increment assumptions of earlier models are 
violated. These results underscore the power of the renewal approach: it 
requires only the first two moments of the cycle-integrated rewards and 
durations, providing a model-independent tool for bounding the precision 
of complex stochastic processes.

We emphasize that the present framework relies on a specific --- though we
believe close to minimal --- structural assumption: the cycles $(N_i,R_i)$
must be i.i.d.\ across renewals, and the auxiliary directional variable
$X_i$ must be i.i.d.\ and independent of $(N_i,R_i)$. 

Run-and-tumble dynamics satisfy this structure only in the free-particle
regime, i.e.\ for force-free motion in a homogeneous, isotropic medium, where an
idealized tumble both resets the internal run statistics and reselects the
direction independently of them. Already for a single run-and-tumble particle
the structure is lost in the presence of an external potential, confinement or
boundaries: the run duration and displacement then depend on position and
orientation, which correlates $X_i$ with $(N_i,R_i)$ and successive cycles with
one another, as is well known from ratchet effects in active matter
\cite{Reichhardt2017}; the same happens in chemotaxis, where run lengths depend on
the orientation relative to the gradient. More generally, this is not a generic
feature of renewal processes: in many systems of interest --- for instance
non-Poissonian enzymatic cycles in which successive turnover times and
displacements remain correlated across cycles, or resetting processes in
which the post-reset orientation is itself correlated with the preceding
cycle --- the required independence is violated, and the bound derived here
does not directly apply. Extending the present construction to such
Markov-modulated or semi-Markov renewal processes, where the $X_i$ (or the
joint law of $(N_i,R_i,X_i)$) form a correlated sequence rather than an
i.i.d.\ one, is a natural and, we believe, tractable direction for future
work.

The primary advantage of the generalized bound derived here is its black-box nature. By focusing on the reward 
magnitude $R$ and the cycle duration $N$ as aggregate variables, we circumvent the need for a fine-grained 
microscopic description of the run phase. As shown in the AR(1) case, the bound remains analytically tractable 
even when internal increments are temporally correlated—a scenario ubiquitous in biological transport but 
notoriously difficult to treat with standard TURs. This suggests that the precision of ``active" transport is 
fundamentally constrained by the discrete statistics of renewal events and the broken symmetry of directional 
choice, rather than the microscopic details of the trajectory.

Furthermore, our results highlight the practical utility of the auxiliary sign variable $X$. In many biophysical 
contexts, the energetic cost is dominated by the ``tumble" or the resetting of directional bias. By relating 
current precision to the entropy production $\bar{s}_X$ of this observable event, we provide a viable 
method for estimating thermodynamic bounds in systems where total entropy production is experimentally 
inaccessible, but the statistics of directional resets are readily measurable. This interpretation 
requires the orientation to be odd under the time reversal of the full dynamics (Sec.~\ref{sec:general_result}); otherwise
$\bar s_X$ quantifies the entropy production of the directional choice in the odd convention only.

Future work may explore the application of this universal bound to experimental data from molecular motors or 
chemotactic bacteria, where internal correlations and resting phases are ubiquitous features of the 
dynamics \cite{Bechinger2016}. Additionally, extending this framework to multi-dimensional renewal 
processes or systems with non-stationary cycle statistics could provide further insights into the universal limits of non-equilibrium transport.
%%%%%%%%%%%%%%%%%%%%%%%%%%%%%%%%%%%%%%%%%%%%%%%%%%%%%%%%%%%
\appendix
\section{Derivation of the Variance Decomposition} \label{app:variance_decomposition}

In this appendix, we detail the derivation of the variance decomposition used in the main text:
$$ \operatorname{Var}(\Delta J - \bar{j}N) = \operatorname{Var}(X) \mathbb{E}[R^2] + m_X^2 \operatorname{Var}(R - \bar{r}N). $$
Let the cycle current be defined as $\Delta J = X R$. 
The sign variable $X \in \{-1, 1\}$, which implies $X^2 = 1$. We denote its expectation as $\mathbb{E}[X] = m_X$.
The variable $X$ is statistically independent of the cycle reward $R$ and the cycle duration $N$.
The mean current is $\bar{j} = \frac{\mathbb{E}[\Delta J]}{\mathbb{E}[N]} = \frac{m_X \mathbb{E}[R]}{\mathbb{E}[N]}$.
The mean run reward rate is defined as $\bar{r} = \frac{\mathbb{E}[R]}{\mathbb{E}[N]}$, which yields the relation $\bar{j} = m_X \bar{r}$.
We first evaluate the variance on the left-hand side, $\operatorname{Var}(XR - \bar{j}N)$. The expected value of this argument is zero:
$$
\begin{aligned} 
\mathbb{E}[XR - \bar{j}N] &= m_X \mathbb{E}[R] - \bar{j} \mathbb{E}[N] \\
&= m_X \mathbb{E}[R] - \left(\frac{m_X \mathbb{E}[R]}{\mathbb{E}[N]}\right) \mathbb{E}[N] = 0. 
\end{aligned}
$$
Since the mean is zero, the variance is the expected value of the squared term:
$$ \operatorname{Var}(XR - \bar{j}N) = \mathbb{E}[(XR - \bar{j}N)^2]. $$
Expanding the square and using $X^2 = 1$ alongside the independence of $X$ from $(R,N)$, we obtain:
$$ 
\begin{aligned}
\mathbb{E}[(XR - \bar{j}N)^2] &= \mathbb{E}[X^2 R^2 - 2\bar{j}XRN + \bar{j}^2 N^2] \\
&= \mathbb{E}[R^2] - 2\bar{j}m_X \mathbb{E}[RN] + \bar{j}^2 \mathbb{E}[N^2].
\end{aligned} 
$$
We now expand the proposed decomposition: $\operatorname{Var}(X) \mathbb{E}[R^2] + m_X^2 \operatorname{Var}(R - \bar{r}N)$.
For the first term, since $X \in \{-1, 1\}$, its variance is $\operatorname{Var}(X) = \mathbb{E}[X^2] - \mathbb{E}[X]^2 = 1 - m_X^2$. Thus:
$$ \operatorname{Var}(X) \mathbb{E}[R^2] = (1 - m_X^2) \mathbb{E}[R^2] = \mathbb{E}[R^2] - m_X^2 \mathbb{E}[R^2]. $$
For the second term, the mean of $(R - \bar{r}N)$ is $\mathbb{E}[R] - \bar{r}\mathbb{E}[N] = 0$. Its variance is therefore:
$$
\begin{aligned}  
\operatorname{Var}(R - \bar{r}N) &= \mathbb{E}[(R - \bar{r}N)^2] \\
&= \mathbb{E}[R^2 - 2\bar{r}RN + \bar{r}^2 N^2] \\
& = \mathbb{E}[R^2] - 2\bar{r}\mathbb{E}[RN] + \bar{r}^2 \mathbb{E}[N^2].
\end{aligned}
$$
Multiplying by $m_X^2$ yields:
$$ m_X^2 \operatorname{Var}(R - \bar{r}N) = m_X^2 \mathbb{E}[R^2] - 2m_X^2 \bar{r} \mathbb{E}[RN] + m_X^2 \bar{r}^2 \mathbb{E}[N^2]. $$
Adding the two expanded terms of the right-hand side, the $m_X^2 \mathbb{E}[R^2]$ terms cancel:
$$ 
\begin{aligned}
\text{RHS} &= \left( \mathbb{E}[R^2] - m_X^2 \mathbb{E}[R^2] \right) \\
&+ \left( m_X^2 \mathbb{E}[R^2] - 2m_X^2 \bar{r} \mathbb{E}[RN] + m_X^2 \bar{r}^2 \mathbb{E}[N^2] \right) \\
&= \mathbb{E}[R^2] - 2m_X(m_X \bar{r})\mathbb{E}[RN] + (m_X \bar{r})^2 \mathbb{E}[N^2].
\end{aligned} 
$$
Substituting the relation $\bar{j} = m_X \bar{r}$ back into the expression gives:
$$ \text{RHS} = \mathbb{E}[R^2] - 2\bar{j}m_X \mathbb{E}[RN] + \bar{j}^2 \mathbb{E}[N^2]. $$
This exactly matches the expanded left-hand side, completing the proof.
%%%%%%%%%%%%%%%%%%%%%%%%%%%%%%%%%%%%%%%%%%%%%%%%%%%%%%%%%%%

\section{PV inequality for a binary variable}\label{app:PV_binary}

Let $X\in\{-1,+1\}$ with $\mathbb{P}(X=+1)=p\in(0,1)$ and $m\equiv m_X=2p-1$, so that
$\operatorname{Var}(X)=1-m^2=4p(1-p)$ and $p/(1-p)=(1+m)/(1-m)$. Then
\[
\bar s_X=(2p-1)\ln\frac{p}{1-p}=m\ln\frac{1+m}{1-m}=2m\,\operatorname{artanh}m .
\]
We show that
\[
\frac{m^2}{1-m^2}\le\frac12\left(e^{\bar s_X}-1\right).
\]
Both sides are even in $m$ and equal zero at $m=0$, so let $0<m<1$. The
inequality is equivalent to $e^{\bar s_X}\ge(1+m^2)/(1-m^2)$, i.e.\ to
\[
g(m)\equiv 2m\,\operatorname{artanh}m-\ln\frac{1+m^2}{1-m^2}\ge0 .
\]
Now $g(0)=0$ and
\[
g'(m)=2\operatorname{artanh}m-\frac{2m}{1+m^2}\ \ge\ 2m-2m=0 ,
\]
since $\operatorname{artanh}m\ge m\ge m/(1+m^2)$ for $0\le m<1$. Hence $g\ge0$ on
$[0,1)$, which proves the inequality. Equality holds only at $m=0$, where both
sides agree to leading order $4\varepsilon^2$ in the small-bias expansion of
Sec.~\ref{sec:small_bias}.
%%%%%%%%%%%%%%%%%%%%%%%%%%%%%%%%%%%%%%%%%%%%%%%%%%%%%%

\end{document}